\documentclass[conference]{IEEEtran}
\input{Qcircuit.tex }
\usepackage{graphicx}
\usepackage{subcaption}
\usepackage{multirow}
\usepackage{amsmath}
\usepackage{amsfonts}
\usepackage{booktabs}
\usepackage{multicol}
\usepackage{multirow}
\usepackage{rotating}
\usepackage{epstopdf}

\IEEEoverridecommandlockouts
\IEEEpubid{	978-1-5386-7466-6/18~\$31.00 \copyright~2018 IEEE }

\begin{document}

\title{ T-count Optimized Quantum Circuits \\ 
	for Bilinear Interpolation}
%
%
\author{\IEEEauthorblockN{Edgard Mu\~{n}oz-Coreas, Himanshu Thapliyal}
\IEEEauthorblockA{ Department of Electrical and Computer Engineering\\
	University of Kentucky, Lexington, KY \\
	Email: hthapliyal@uky.edu}                %
}


%

%

%
\ifCLASSINFOpdf
\else
\fi

\maketitle



%
\IEEEpeerreviewmaketitle

\begin{abstract}

Quantum circuits for basic image processing functions such as bilinear interpolation are required to implement image processing algorithms on quantum computers.  In this work, we propose quantum circuits for the bilinear interpolation of NEQR encoded images based on Clifford+T gates.  Quantum circuits for the scale up operation and scale down operation are illustrated.  The proposed quantum circuits are based on quantum Clifford+T gates and are optimized for T-count. Quantum circuits based on Clifford+T gates can be made fault tolerant but the T gate is very costly to implement.  As a result, reducing T-count is an important optimization goal.  The proposed quantum bilinear interpolation circuits are based on (i) a quantum adder, (ii) a proposed quantum subtractor, and (iii) a quantum multiplication circuit. Further, both designs are compared and shown to be superior to existing work in terms of T-count.  The proposed quantum bilinear interpolation circuits for the scale down operation and for the scale up operation each have a 92.52\% improvement in terms of T-count compared to the existing work. 

\end{abstract}

\section{Introduction}
Quantum computing has promising applications in number theory, encryption, search, scientific computation and image processing. For example quantum algorithms have been proposed for image orientation problems, image pattern recognition and image template matching \cite{Beach2004motivation} \cite{Caraiman2012motivation}. Quantum image representations and image manipulation are required in the quantum circuit implementations of image processing quantum algorithms.  Thus, researchers have proposed quantum image representations such as the Flexible Representation of Quantum Images (FRQI) \cite{Le2011FRQI} and the Novel Enhanced Quantum Representation (NEQR) \cite{Zhang2013NEQR}. Further, quantum circuits for image operations such as translation, geometric transformation and bilinear interpolation have also been developed \cite{Zhou2017bilinear} \cite{Jiang2015interpolation} \cite{YanFei2017rotation}.  If these quantum circuits are based on Clifford+T gates, they can be made fault tolerant with error correcting codes permitting reliable and scalable quantum computation \cite{Maslov2013cliffordT} \cite{PalerIOP} \cite{HowardMark2017Tgates} \cite{Zhou-T} \cite{Kiteav2005faulttolerant}.  The Clifford+T gate family is illustrated in \cite{Maslov2013cliffordT}.  The T gate is very costly to implement compared to the Clifford gates making T-count an important optimization goal \cite{PalerIOP} \cite{Maslov2013cliffordT} \cite{HowardMark2017Tgates} \cite{Zhou-T} \cite{Kiteav2005faulttolerant}.  

The design of quantum circuits for operations such as interpolation for NEQR images has been addressed in the literature \cite{Zhou2017bilinear}.  The existing bilinear interpolation quantum circuits are based on (i) a quantum subtractor, (ii) a quantum addition circuit, (iii) a quantum multiplication circuit, and (iv) a quantum division circuit.  The work in \cite{Zhou2017bilinear} also proposes a color information retrieval scheme based on quantum oracles operating as lookup tables.  While an interesting design, the bilinear interpolation circuits in \cite{Zhou2017bilinear} suffers from significant T gate cost because the design is based on quantum arithmetic circuits that have high T-count.  Further, the circuits suffer from added overhead due to extra quantum arithmetic operations which can be minimized.  

To overcome the limitations of the existing designs, this work presents quantum circuits for bilinear interpolation of NEQR encoded images based on Clifford+T gates.  Quantum circuits for the scale up operation and scale down operation are illustrated.  The proposed quantum bilinear interpolation circuits are based on: (i) a quantum adder, (ii) a proposed quantum subtractor and (iii) a quantum multiplication circuit.  The proposed quantum bilinear interpolation circuits do not require a quantum division circuit.  The proposed quantum bilinear interpolation circuits are compared and shown to be superior to the existing work in terms of T-count and number of arithmetic units used.  This paper is organized as follows: Section \ref{bi-background} discusses the Clifford+T gate set and discusses the novel enhanced quantum representation (NEQR).  Section \ref{bi-components} illustrates the quantum adder, proposed quantum subtractor and quantum multiplication circuit used in the proposed bilinear interpolation circuits.  In Section \ref{design-bi-down}, the design of the proposed quantum bilinear interpolation circuit for the scale down operation is presented and compared to the existing work.  Lastly, in Section \ref{design-bi-up}, the design of the proposed quantum bilinear interpolation circuit for the scale up operation is presented and compared to the existing work.

\IEEEpubidadjcol

\section{Background}
\label{bi-background}

\subsection{Quantum Gates}

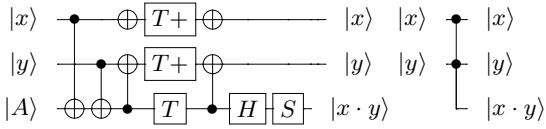
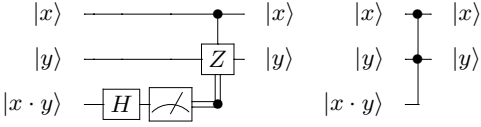
\begin{figure}
\centering
\begin{subfigure}[hb]{3in}
\small
\[
\Qcircuit @C=0.3em @R=0.5em @!R{
\lstick{\ket{x}} & & \ctrl{2} &\qw &\targ & \gate{T+} & \targ & \qw & \qw & \qw & \qw &\qw &\rstick{\ket{x}} & & & & & & & & & & & & & & & \lstick{\ket{x}} & & \ctrl{1} & \qw & \rstick{\ket{x}}\\
\lstick{\ket{y}} & & \qw &\ctrl{1} &\targ & \gate{T+} & \targ & \qw & \qw & \qw & \qw &\qw &\rstick{\ket{y}}  & & & & & & & &   & & & & & & & \lstick{\ket{y}} & & \ctrl{1} & \qw & \rstick{\ket{y}} \\
\lstick{\ket{A}} & & \targ &\targ &\ctrl{-1} & \gate{T} & \ctrl{-1} & \gate{H} & \gate{S} &\qw &\rstick{\ket{x \cdot y}} & & & & & & & & & & & & &  & & & & & & \qwx &\qw &\rstick{\ket{x \cdot y}} \\  
}
\]
\caption{The temporary logical-AND gate and its Clifford+T gate implementation.  This Clifford+T gate implementation of the temporary logical-AND gate has a T-count of $4$. $\ket{A}$ is an ancillae in the state $\frac{1}{\sqrt{2}}(\ket{0} + e^{\frac{i \cdot \pi}{4}}\ket{1})$.}
\label{bi-logical-AND}
\end{subfigure} 
\\
\begin{subfigure}[hb]{3in}

\end{subfigure}
\\
\begin{subfigure}[hb]{3in}

\end{subfigure}
\\
\begin{subfigure}[hb]{3in}
\small
\[
\Qcircuit @C=0.4em @R=0.5em @!R{
\lstick{\ket{x}} & & \qw &\qw & \qw & \ctrl{1} &\qw & \rstick{\ket{x}} & & & & & & & & & & & & & & & \lstick{\ket{x}} & & \ctrl{1} & \qw & \rstick{\ket{x}}\\
\lstick{\ket{y}} & & \qw & \qw & \qw & \gate{Z} &\qw & \rstick{\ket{y}} & & & & & & & & & & & & & & & \lstick{\ket{y}} & & \ctrl{1} & \qw & \rstick{\ket{y}} \\
\lstick{\ket{x \cdot y}} & & \qw &  \gate{H} & \meter & \control \cw \cwx & & & & & & & & & & & & & & & & & \lstick{\ket{x \cdot y}} &  & \qw & & & \\
}
\]
\caption{The uncomputation gate and its Clifford+T gate implementation.  This Clifford+T gate implementation of the uncomputation gate has a T-count of $0$. }
\label{bi-uncompute}
\end{subfigure}
\caption{The quantum gates presented in \cite{Gidney20184TgateToffolibuild} used in this work.  These gates are derived from the designs in \cite{Cody2012toffoligatebuild}.  Quantum gate and graphical representations are shown.} 
\end{figure}

Fault tolerant implementation of quantum circuits is of interest to researchers because physical quantum computers are prone to noise errors \cite{Maslov2013cliffordT} \cite{PalerIOP} \cite{Kiteav2005faulttolerant}.  Recently, researchers have implemented quantum logic gates and circuits with the Clifford+T gate set because they can be made fault tolerant \cite{Maslov2013cliffordT} \cite{PalerIOP} \cite{HowardMark2017Tgates} \cite{Zhou-T} \cite{Kiteav2005faulttolerant}.  The set of gates that make up the Clifford+T gate family is illustrated in \cite{Maslov2013cliffordT}.  The bilinear interpolation circuits proposed in this work are composed of the NOT gate, the Feynman (CNOT) gate, the temporary logical-
AND gate and uncomputation gate.  The CNOT gate and the NOT gate are in the set of gates that make up the Clifford+T gate family \cite{Maslov2013cliffordT}.  The temporary logical-AND gate and uncomputation gate must be constructed from Clifford+T gates and are presented in \cite{Gidney20184TgateToffolibuild}.  These gates are based on the designs in \cite{Cody2012toffoligatebuild}.  The temporary logical-AND gate is a 3 input, 3 output logic gate and has the mapping $A,B, \frac{1}{\sqrt{2}}(\ket{0} + e^{\frac{i \cdot \pi}{4}}\ket{1}) $ to $A,B,A \cdot B$.  The Clifford+T gate implementation of the temporary logical-AND gate  is shown in Figure \ref{bi-logical-AND}.  The input labeled $\ket{A}$ in Figure \ref{bi-logical-AND} is an ancillae in the state shown in expression \ref{bi-ancillary}:

\begin{equation}
 \frac{1}{\sqrt{2}}(\ket{0} + e^{\frac{i \cdot \pi}{4}}\ket{1})
\label{bi-ancillary} 
 \end{equation}
 
 By using the ancillae set to $\ket{A}$, the temporary logical-AND gate designed in \cite{Cody2012toffoligatebuild} can be realized on three qubits instead of four \cite{Gidney20184TgateToffolibuild}.
 
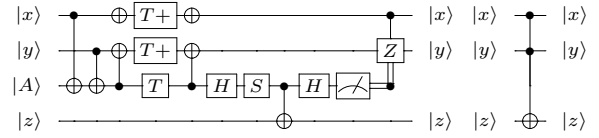
\begin{figure}[tbhp]
\scriptsize
\[
\Qcircuit @C=0.4em @R=0.5em @!R{
\lstick{\ket{x}} & & \ctrl{2} &\qw &\targ & \gate{T+} & \targ & \qw & \qw & \qw & \qw & \qw & \ctrl{1} &\qw & \rstick{\ket{x}} & & & & & & & & & & & & \lstick{\ket{x}} & & \ctrl{2} & \qw & \rstick{\ket{x}}\\
\lstick{\ket{y}} & & \qw &\ctrl{1} &\targ & \gate{T+} & \targ & \qw & \qw & \qw & \qw & \qw & \gate{Z} &\qw & \rstick{\ket{y}} & & & & & & & & & & & & \lstick{\ket{y}} & & \ctrl{2} & \qw & \rstick{\ket{y}} \\
\lstick{\ket{A}} & & \targ &\targ &\ctrl{-1} & \gate{T} & \ctrl{-1} & \gate{H} & \gate{S} & \ctrl{1} &  \gate{H} & \meter & \control \cw \cwx & & & & & & & & & & & & & & & & \\
\lstick{\ket{z}} & & \qw & \qw & \qw & \qw & \qw & \qw & \qw & \targ & \qw &\qw &\qw &\qw &\rstick{\ket{z}} & & & & & & & & & & & & \lstick{\ket{z}} & & \targ & \qw & \rstick{\ket{z}}  
}
\]
\caption{The Toffoli gate and its Clifford+T gate implementation \cite{Cody2012toffoligatebuild}.  This Clifford+T gate implementation of the Toffoli gate has a T-count of $4$. $\ket{A}$ is an ancillae in the state $\frac{1}{\sqrt{2}}(\ket{0} + e^{\frac{i \cdot \pi}{4}}\ket{1})$.}
\label{Bi-toffoli-implementation}
\end{figure}

   The uncomputation gate is a 3 input, 3 output logic gate and has the mapping $A,B,A \cdot B$  to $A,B,0$.  The third input 
 qubit can be restored to an ancillae after measurement for use in later computation.  The Clifford+T gate implementation of the uncomputation gate is shown in Figure \ref{bi-uncompute}.  A Toffoli gate can be realized from the temporary logical-AND gate and uncomputation gate \cite{Cody2012toffoligatebuild}.  The Toffoli gate implementation is shown in Figure \ref{Bi-toffoli-implementation}.

Evaluating quantum circuit performance in terms of the number of T gates (T-count) is of interest because the fault tolerant implementation of the T gate is significantly more costly than the fault tolerant implementation costs of the other Clifford+T gates \cite{PalerIOP} \cite{Maslov2013cliffordT} \cite{HowardMark2017Tgates} \cite{Zhou-T} \cite{Kiteav2005faulttolerant}. By using the logical-AND gate and the uncomputation gate to realize quantum arithmetic circuits, our proposed bilinear interpolation circuits reduce the number of T gates used.

\subsection{Novel Enhanced Quantum Representation (NEQR)}

An image must be represented as qubits.  Researchers have proposed several methods to represent both color and greyscale images on a quantum computer \cite{Zhang2013NEQR} \cite{Le2011FRQI}.  In this work, we use the Novel Enhanced Quantum Representation (NEQR) presented in \cite{Zhang2013NEQR}.  NEQR is a means to represent greyscale images on a quantum machine.  For a given image, each pixel is represented with expression \ref{bi-eq:2}:      

\begin{equation}
	\ket{Y} \ket{X} \ket{C}
	\label{bi-eq:2} 
\end{equation}

Where $\ket{Y}$ and $\ket{X}$ are quantum registers that contain the $x$ and $y$ coordinates of the image and $\ket{C}$ contains the greyscale color of the pixel.  Quantum registers $\ket{Y}$ and $\ket{X}$ are of size $n$ and the color information quantum register $\ket{C}$ is of size $q$ to store the needed color information \cite{Zhang2013NEQR} \cite{Zhou2017bilinear}.  NEQR improves the existing encoding scheme FRQI (Flexible Representation of Quantum Images) because the image color information is represented as multiple qubits in the computational basis as opposed to a single qubit in superpositiion.  NEQR benefits from faster image preparation, accurate color measurement and easier implementation of quantum image processing circuits compared to FRQI \cite{Zhang2013NEQR}.

\section{Design of Quantum Circuits Used In Proposed Bilinear Interpolation Circuits}
\label{bi-components}

The proposed quantum bilinear interpolation circuits are based on: (i) a quantum adder, (ii) a proposed quantum subtractor and (iii) a quantum multiplication circuit.
The circuit designs of the quantum adder, quantum subtractor and quantum multiplier are discussed below:

\begin{itemize}

\item \textbf{Quantum Adder:} We use the quantum ripple carry adder presented in \cite{Gidney20184TgateToffolibuild} in this work.  The quantum adder is based on the ripple carry adder design in \cite{Draper}.  The quantum circuit takes two $n$ bit inputs $A$ and $B$.  At the end of computation, the input $A$ emerges unchanged and the input $B$ is transformed to the sum $B + A$.  The quantum adder saves T gates by using the logical-AND gate and the uncomputation gate implementations shown in Section \ref{bi-background}.  The design of the quantum addition circuit is illustrated in \cite{Gidney20184TgateToffolibuild}.

\item \textbf{Proposed Quantum Subtractor:}  We propose a quantum subtraction circuit in this work.  The proposed quantum subtraction circuit is shown in Figure \ref{bl: new-subtractor}.  

\begin{figure}[tbhp]
\flushleft
\small
\[
\Qcircuit @C=0.4em @R=0.5em @!R{
\lstick{\ket{A_0}} &\qw & \ctrl{2} & \qw & \qw  &\qw &\qw &\qw &\qw &\qw &\qw &\qw &\qw &\qw  &\qw &\qw &\ctrl{2} &\ctrl{1} &\qw &\qw &\rstick{\ket{A_0}}\\
\lstick{\ket{B_0}} &\targ & \ctrl{1} &\qw  & \qw &\qw  &\qw  &\qw &\qw &\qw &\qw &\qw &\qw &\qw &\qw  &\qw &\ctrl{1} &\targ &\targ &\qw &\rstick{\ket{S_0}}\\
& & \qwx & \ctrl{2} &\qw &\ctrl{3} &\qw &\qw  &\qw &\qw &\qw &\qw &\qw &\ctrl{3} &\qw &\ctrl{1} &\qwx \qw & & \\
\lstick{\ket{A_1}} &\qw & \qw & \targ & \ctrl{2} &\qw &\qw &\qw &\qw &\qw &\qw &\qw &\qw &\qw &\ctrl{2} &\targ &\qw  &\ctrl{1} &\qw &\qw &\rstick{\ket{A_1}} \\
\lstick{\ket{B_1}} &\targ & \qw & \targ & \ctrl{1} &\qw &\qw &\qw &\qw \qw &\qw &\qw &\qw &\qw  &\qw &\ctrl{1} &\qw &\qw &\targ &\targ &\qw &\rstick{\ket{S_1}}\\
& & & &\qwx  &\targ &\ctrl{2} &\qw &\ctrl{3} &\qw &\ctrl{3} &\qw &\ctrl{1} &\targ &\qw \qwx & & & \\
\lstick{\ket{A_2}} &\qw &\qw  & \qw & \qw & \qw &\targ &\ctrl{2} &\qw &\qw  &\qw &\ctrl{2} &\targ &\qw &\qw &\qw &\qw  &\ctrl{1} &\qw &\qw &\rstick{\ket{A_2}}\\
\lstick{\ket{B_2}} &\targ &\qw  & \qw & \qw & \qw &\targ &\ctrl{1} &\qw &\qw &\qw   &\ctrl{1} &\qw &\qw &\qw &\qw &\qw &\targ &\targ &\qw &\rstick{\ket{S_2}}\\
& & & & & & &\qwx &\targ &\ctrl{2} &\targ &\qw \qwx & & & & & & & & \\
\lstick{\ket{A_3}} &\qw &\qw  & \qw & \qw & \qw &\qw  &\qw &\qw &\qw &\qw &\qw  &\qw &\qw &\qw &\qw &\qw &\ctrl{1} &\qw &\qw &\rstick{\ket{A_3}}\\
\lstick{\ket{B_3}} &\targ &\qw  & \qw & \qw & \qw &\qw &\qw &\qw &\targ &\qw &\qw  &\qw &\qw &\qw &\qw &\qw &\targ &\targ &\qw &\rstick{\ket{S_3}}\\
}
\]
\caption{Proposed quantum subtraction circuit for four qubit operands.}
\label{bl: new-subtractor}
\end{figure}
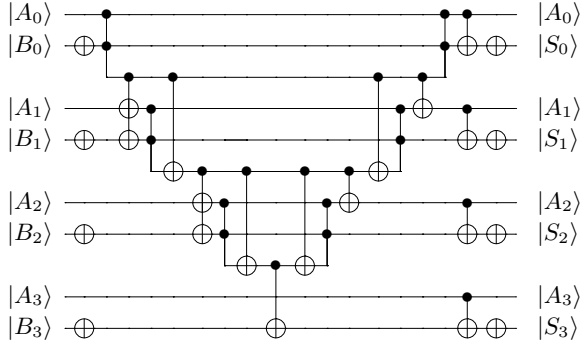

To save T gates, the proposed quantum subtraction circuit is based on the quantum ripple carry adder presented in \cite{Gidney20184TgateToffolibuild}.  To get the ripple carry adder to perform subtraction, we use the design approach presented in \cite{Thapliyal2016addsub}.  As shown in Figure \ref{bl: new-subtractor}, the input $B$ is complemented before being applied to the quantum ripple carry adder.  Thus, the ripple carry adder calculates $\bar{B}+A$.  Afterward, the qubits that originally held $B$ are complemented again.  As a result, the quantum subtractor calculates $(\overline{\bar{B}+A})$ at the end of computation.  $(\overline{\bar{B}+A})$ is equivalent to $B-A$ \cite{Thapliyal2016addsub}.  While the example in Figure \ref{bl: new-subtractor} is sized for 4 qubit operands, the proposed quantum subtraction circuit design can be extended to any operand size.  

\item \textbf{Quantum Multiplier:} We use the quantum integer multiplication circuit presented in \cite{edgard2017multiplier} in this work.  The quantum multiplication circuit takes two $n$ bit inputs $a$ and $b$.  At the end of computation, the inputs $a$ and $b$ emerges unchanged and the product $b \cdot a$ is generated on ancillae.  To save T gates, we use the conditional adder presented in \cite{Gidney20184TgateToffolibuild} in the multiplication circuit.  The quantum conditional adder circuit described in \cite{Gidney20184TgateToffolibuild} takes two $n$ bit inputs $a$ and $b$ and a $1$ bit input $control$.  When the $control = 1$, the circuit calculates $b + a$ and when $control = 0$ the circuit performs no computation.    

\end{itemize}

\section{Design of the Proposed Bilinear Interpolation Circuit for the Scale Down Operation}
\label{design-bi-down}

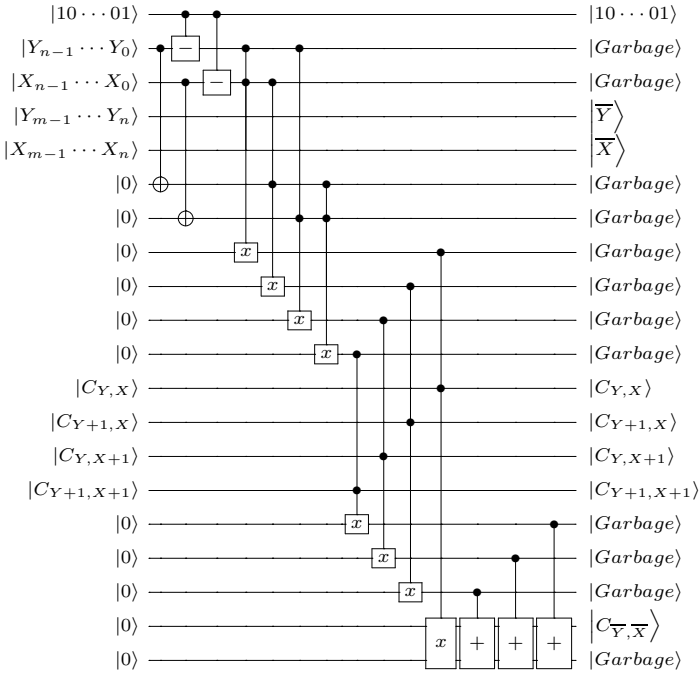
\begin{figure}[tbhp]
\flushleft
\scriptsize 
\[
\Qcircuit @C=0.2em @R=0.5em @!R{
\lstick{\ket{10 \cdots 01}} & \qw &\ctrl{1} &\ctrl{2} & \qw & \qw & \qw & \qw & \qw & \qw & \qw & \qw & \qw & \qw & \qw & \qw & \qw & \rstick{\ket{10 \cdots 01}}\\
\lstick{\ket{Y_{n-1} \cdots Y_0} } & \ctrl{4} &\gate{-} & \qw &\ctrl{3} &\qw &\ctrl{5}  & \qw & \qw & \qw & \qw & \qw & \qw & \qw & \qw & \qw & \qw& \rstick{\ket{Garbage} }\\
\lstick{\ket{X_{n-1} \cdots X_0} } & \qw & \ctrl{4} &\gate{-} &\ctrl{5} & \ctrl{3} &\qw & \qw & \qw & \qw & \qw & \qw & \qw & \qw & \qw & \qw & \qw& \rstick{\ket{Garbage} }\\
\lstick{\ket{Y_{m-1} \cdots Y_n} } & \qw & \qw & \qw & \qw &\qw &\qw &\qw & \qw & \qw & \qw & \qw & \qw & \qw & \qw & \qw & \qw & \rstick{\ket{\overline{Y}}}\\
\lstick{\ket{X_{m-1} \cdots X_n} } & \qw & \qw & \qw & \qw & \qw &\qw &\qw & \qw & \qw & \qw & \qw & \qw & \qw & \qw & \qw & \qw& \rstick{\ket{\overline{X}}}\\
\lstick{\ket{0}} & \targ & \qw & \qw & \qw & \ctrl{3} & \qw & \ctrl{3} & \qw & \qw & \qw & \qw & \qw & \qw & \qw & \qw & \qw & \rstick{\ket{Garbage} }\\
\lstick{\ket{0}} & \qw & \targ & \qw & \qw & \qw & \ctrl{3} & \ctrl{4} & \qw & \qw & \qw & \qw & \qw & \qw & \qw & \qw & \qw& \rstick{\ket{Garbage} }\\
\lstick{\ket{0}} & \qw & \qw & \qw &\gate{x} & \qw & \qw & \qw & \qw & \qw & \qw & \qw &\ctrl{4} & \qw & \qw & \qw & \qw &\rstick{\ket{Garbage} }\\
\lstick{\ket{0}} & \qw & \qw & \qw &\qw &\gate{x} & \qw & \qw & \qw & \qw & \qw &\ctrl{4} & \qw & \qw & \qw & \qw & \qw &\rstick{\ket{Garbage} }\\
\lstick{\ket{0}} & \qw & \qw & \qw & \qw & \qw &\gate{x} & \qw & \qw & \qw &\ctrl{4} & \qw & \qw & \qw & \qw & \qw & \qw &\rstick{\ket{Garbage} }\\
\lstick{\ket{0}} & \qw & \qw & \qw & \qw & \qw & \qw &\gate{x} & \qw &\ctrl{4} & \qw & \qw & \qw & \qw & \qw & \qw & \qw &\rstick{\ket{Garbage} }\\
\lstick{\ket{C_{Y,X}}} & \qw & \qw & \qw & \qw & \qw & \qw & \qw & \qw & \qw & \qw & \qw &\ctrl{7} & \qw & \qw & \qw & \qw& \rstick{\ket{C_{Y,X}}}\\
\lstick{\ket{C_{Y+1,X}}} & \qw & \qw & \qw & \qw & \qw & \qw & \qw & \qw & \qw & \qw &\ctrl{5} & \qw & \qw & \qw & \qw  & \qw& \rstick{\ket{C_{Y+1,X}}}\\
\lstick{\ket{C_{Y,X+1}}} & \qw & \qw & \qw & \qw & \qw & \qw & \qw & \qw & \qw & \ctrl{3} & \qw & \qw & \qw & \qw & \qw  & \qw& \rstick{\ket{C_{Y,X+1}}}\\
\lstick{\ket{C_{Y+1,X+1}}} & \qw & \qw & \qw & \qw & \qw & \qw & \qw & \qw & \ctrl{1} & \qw  & \qw & \qw & \qw & \qw & \qw & \qw& \rstick{\ket{C_{Y+1,X+1}}}\\
\lstick{\ket{0}} & \qw & \qw & \qw & \qw & \qw & \qw & \qw & \qw & \gate{x} & \qw & \qw & \qw & \qw & \qw &\ctrl{3} & \qw& \rstick{\ket{Garbage} }\\
\lstick{\ket{0}} & \qw & \qw & \qw & \qw & \qw & \qw & \qw & \qw & \qw &\gate{x} & \qw & \qw & \qw &\ctrl{2} & \qw & \qw &\rstick{\ket{Garbage} }\\
\lstick{\ket{0}} & \qw & \qw & \qw & \qw & \qw & \qw & \qw & \qw & \qw &\qw & \gate{x} & \qw &\ctrl{1} & \qw & \qw & \qw &\rstick{\ket{Garbage} }\\
\lstick{\ket{0}} & \qw & \qw & \qw & \qw & \qw & \qw & \qw & \qw & \qw & \qw &\qw & \multigate{1}{x} & \multigate{1}{+} & \multigate{1}{+} & \multigate{1}{+} & \qw &\rstick{\ket{C_{\overline{Y},\overline{X}}}}\\
\lstick{\ket{0}} & \qw & \qw & \qw & \qw & \qw & \qw & \qw & \qw & \qw & \qw &\qw & \ghost{x} & \ghost{+} & \ghost{+} & \ghost{+} & \qw &\rstick{\ket{Garbage} }\\
}
\]
\caption{Proposed quantum bilinear interpolation circuit for the scale down operation.  The image is scaled down by a value $n$. }
\label{bilinear-scale-down-image}
\end{figure}

The proposed quantum circuit for bilinear interpolation is shown in Figure \ref{bilinear-scale-down-image} for the case of a scaling down an image by an integer value $n$.  Scaling down an image by an integer value $n$ results in reducing the original $y$ and $x$ positions of each pixel by $2^n$.  Inputs to the proposed quantum circuit is the original pixel positions $y$ and $x$ stored in $m$ bit quantum registers $\ket{Y}, \ket{X}$, respectively, along with the corresponding original pixel color information $C_{Y,X}$ stored in quantum register $\ket{C_{Y,X}}$.  Inputs also include the pixel color information for adjacent pixels at locations ($y+1,x$), ($y,x+1$) and ($y+1,x+1$).  The color information for these adjacent pixels are stored in quantum registers $\ket{C_{Y+1,X}}$, $\ket{C_{Y,X+1}}$ and $\ket{C_{Y+1,X+1}}$ respectively.  At the end of computation, the circuitry returns the position information for the scaled pixel in quantum registers $\ket{\overline{Y}}$ and $\ket{\overline{X}}$ as well as the corresponding color of the scaled pixel in quantum register $\ket{C_{\overline{Y},\overline{X}}}$.

Our proposed design calculates the position information for the scaled pixel without quantum gates.  To reduce an image by a value $n$, the original position value is divided by $2^n$.  We accomplish this division by assigning the values at locations $\ket{Y_{m-1}}$ through $\ket{Y_{n}}$ of quantum register $\ket{Y}$ and locations $\ket{X_{m-1}}$ through $\ket{X_{n}}$ of quantum register $\ket{X}$ to the output position registers $\ket{\overline{Y}}$ and $\ket{\overline{X}}$, respectively. Thus, we eliminate the need to use a division circuit.  

To calculate the color information for the scaled pixel, our proposed quantum circuit must perform the calculation shown in expression \ref{blabs:1}:

{\small 
\begin{equation}
\left[ \begin{matrix}
(2^n - (\overline{Y} \cdot 2^n - Y) \cdot (2^n - (\overline{X} \cdot 2^n - X) \cdot C_{Y,X} + \\
(\overline{Y} \cdot 2^n - Y) \cdot (2^n -(\overline{X} \cdot 2^n - X) \cdot C_{Y+1,X} + \\
(2^n - (\overline{Y} \cdot 2^n - Y) \cdot (\overline{X} \cdot 2^n - X) \cdot C_{Y,X+1} + \\
(\overline{Y} \cdot 2^n - Y) \cdot (\overline{X} \cdot 2^n - X) \cdot C_{Y+1,X+1} \\
\end{matrix} \right] \div 2^{2 \cdot n}
\label{blabs:1}
\end{equation}
}

Where $(\overline{Y} \cdot 2^n - Y)$ corresponds to locations $\ket{Y_{n-1}}$ through $\ket{Y_0}$ of quantum register $\ket{Y}$ and $(\overline{X} \cdot 2^n - X)$ corresponds to locations $\ket{X_{n-1}}$ through $\ket{X_0}$ of quantum register $\ket{X}$. 

To perform this computation, the proposed bilinear interpolation circuit requires a quantum multiplication circuit, a quantum addition circuit and a quantum subtraction circuit.  Our circuit computes equation \ref{blabs:1} by executing the following algorithm:

\begin{itemize}

\item Step 1: Copy $(\overline{Y} \cdot 2^n - Y)$ and $(\overline{X} \cdot 2^n - X)$ to ancillae with CNOT gates.

\item Step 2: Calculate $2^n - (\overline{Y} \cdot 2^n - Y)$ and $2^n - (\overline{X} \cdot 2^n - X)$ with quantum subtraction circuits.  

\item Step 3: Calculate the products $(2^n - (\overline{Y} \cdot 2^n - Y) \cdot (2^n - (\overline{X} \cdot 2^n - X)$, $(\overline{Y} \cdot 2^n - Y) \cdot (2^n -(\overline{X} \cdot 2^n - X)$, $(2^n - (\overline{Y} \cdot 2^n - Y) \cdot (\overline{X} \cdot 2^n - X)$ and $(2^n - (\overline{Y} \cdot 2^n - Y) \cdot (\overline{X} \cdot 2^n - X)$ with quantum multiplication circuits.

\item Step 4: Calculate the product terms $(2^n - (\overline{Y} \cdot 2^n - Y) \cdot (2^n - (\overline{X} \cdot 2^n - X) \cdot C_{Y,X}$, $(\overline{Y} \cdot 2^n - Y) \cdot (2^n -(\overline{X} \cdot 2^n - X) \cdot C_{Y+1,X}$, $(2^n - (\overline{Y} \cdot 2^n - Y) \cdot (\overline{X} \cdot 2^n - X) \cdot C_{Y,X+1}$ and $(2^n - (\overline{Y} \cdot 2^n - Y) \cdot (\overline{X} \cdot 2^n - X) \cdot C_{Y+1,X+1}$ with quantum multiplication circuits.

\item Step 5: Complete the calculation of equation \ref{blabs:1} with quantum addition circuits. 

\end {itemize}

The values at locations $2 \cdot n - 1$ through $0$ of the quantum register with the result of Step 5 are not a part of the new pixel's color information $\left( C_{\overline{Y},\overline{X}} \right) $.  By not assigning these locations to the quantum register containing the new pixel's color information $\ket{C_{\overline{Y},\overline{X}}}$ we eliminate the need to use quantum division circuits.  The remaining locations are the new pixel's color information $\left( C_{\overline{Y},\overline{X}} \right) $ and will be assigned to the quantum register containing the new pixel's color information $\ket{C_{\overline{Y},\overline{X}}}$.

\subsection{Cost Analysis of the Proposed Bilinear Interpolation Circuit for the Scale Down Operation}
\label{cost-bi-down}

\begin{table}[tbhp]
\centering
\caption{Comparison of the Bilinear Interpolation Circuits for the Scale Down Operation in Terms of Number of Functional Blocks Used}
\label{bilinear-table:3}
\begin{tabular}{lccc}
\\ \midrule
Component & & 1 & proposed \\ \toprule
Adder & &$3$ & $3$ \\
Subtractor & & $4$ & $2$ \\
Divider & & $2$ & $0$* \\
Multiplier & & $8 $ & $8$ \\ \bottomrule
\multicolumn{4}{l}{1 is the design in \cite{Zhou2017bilinear}} \\
\multicolumn{4}{l}{* Our proposed design does not require a divider.} \\ \midrule
\end{tabular}
\end{table}

\begin{table*}[tbhp]
\centering
\caption{T Gate Comparison of the Bilinear Interpolation Circuits for the Scale Down Operation}
\label{bilinear-table:2}
\begin{tabular}{lccc}
\\ \midrule
&  & T-count & \% impr. w.r.t. 1 \\ \toprule
1 & & $856 \cdot n^2 + 196 \cdot n - 98 + 8 \cdot \sum_{i = 1}^{log_2(n)} \frac{n}{2^i} \cdot \left( 14 \cdot \left( n + i - 2^{i-1} \right) -14 \right)$ & - \\ 
Proposed & & $64 \cdot n^2 - 12 \cdot n - 8$ & $ \approx 92.52 \%$ \\ \bottomrule
\multicolumn{4}{l}{1 is the design in \cite{Zhou2017bilinear}} \\
\end{tabular}
\end{table*}

 Table \ref{bilinear-table:3} shows the comparison between the proposed quantum circuit for bilinear interpolation and the existing work in terms of total number of arithmetic operations.  Through careful layout of functional blocks in our proposed quantum circuit for bilinear interpolation we remove 2 quantum subtraction circuits from our design.  By not assigning locations $2 \cdot n - 1$ through $0$ of the quantum register containing the result of equation \ref{blabs:1} to the quantum register containing the new pixel's color information $\ket{C_{\overline{Y},\overline{X}}}$ we eliminate the need to use quantum division circuits.        
 
  Comparison of the T-count between the proposed quantum bilinear interpolation circuit for the scale down operation and the existing work in \cite{Zhou2017bilinear} is shown in Table \ref{bilinear-table:2}.  The design in \cite{Zhou2017bilinear} has a T-count of order $ \approx \mathcal{O}(n^2)$. The proposed design's T-count is of order $\mathcal{O}(n^2)$.  To calculate the T-count of the existing design in \cite{Zhou2017bilinear}, we determined the T-counts for the quantum circuits used in the design.  We determined that to calculate equation \ref{blabs:1}, the design in \cite{Zhou2017bilinear} uses a quantum addition circuit with a T-count of $28 \cdot n - 14$, a quantum subtraction circuit with a T-count of $28 \cdot n - 14$, a quantum multiplication circuit with T-count  of $7 \cdot n^2 + \sum_{i = 1}^{log_2(n)} \frac{n}{2^i} \cdot \left( 14 \cdot \left( n + i - 2^{i-1} \right) -14 \right) $ and a quantum division circuit with a T-count of $\approx 400 \cdot n^2$.  We compute the T-count for the proposed work and the existing design by multiplying the T-count for each quantum functional block by the number of times it is used and then sum the result.  Table \ref{bilinear-table:3} shows the number of functional blocks used by the design in \cite{Zhou2017bilinear}.  

Table \ref{bilinear-table:2} shows that the proposed quantum circuit for bilinear interpolation have an improvement ratio of $ \approx 92.52 \%$ in terms of T-count when performing the scale down operation.  Our proposed quantum bilinear interpolation circuits save T gates by (i) using T gate efficient functional blocks and (ii) reducing the number of required functional blocks to calculate the new pixel's color information (see equation \ref{blabs:1}).  Our proposed designs are based on the quantum arithmetic circuits illustrated in Section \ref{bi-components}.  As a result, our proposed bilinear interpolation quantum circuit is based on a quantum addition circuit with a T-count of $4 \cdot n$, a novel quantum subtraction circuit with a T-count of $4 \cdot n-4$ and a quantum multiplication circuit with a T-count of $8 \cdot n^2 - 4 \cdot n$.  

\section{Design of the Proposed Bilinear Interpolation Circuits for the Scale Up Operation}
\label{design-bi-up}

\begin{figure}[tbhp]
	\flushleft
	\scriptsize
	\[
	\Qcircuit @C=0.2em @R=0.5em @!R{
		\lstick{\ket{10 \cdots 01}} & \qw &\ctrl{3} &\ctrl{4} & \qw & \qw & \qw & \qw & \qw & \qw & \qw & \qw & \qw & \qw & \qw & \qw & \qw & \rstick{\ket{10 \cdots 01}}\\
		\lstick{\ket{0}} & \qw &\qw & \qw &\qw &\qw &\qw  & \qw & \qw & \qw & \qw & \qw & \qw & \qw & \qw & \qw & \qw& \rstick{\ket{\overline{Y_{n-1:0}} }}\\
		\lstick{\ket{0 }} & \qw & \qw &\qw &\qw & \qw &\qw & \qw & \qw & \qw & \qw & \qw & \qw & \qw & \qw & \qw & \qw& \rstick{\ket{\overline{X_{n-1:0}} }}\\
		\lstick{\ket{Y} } & \ctrl{2} & \gate{-} & \qw & \ctrl{3} &\qw &\ctrl{4} &\qw & \qw & \qw & \qw & \qw & \qw & \qw & \qw & \qw & \qw & \rstick{\ket{Garbage} }\\
		\lstick{\ket{X} } & \qw & \ctrl{2} & \gate{-} & \ctrl{3} & \ctrl{1} &\qw &\qw & \qw & \qw & \qw & \qw & \qw & \qw & \qw & \qw & \qw& \rstick{\ket{Garbage} }\\
		\lstick{\ket{0}} & \targ & \qw & \qw & \qw & \ctrl{3} & \qw & \ctrl{3} & \qw & \qw & \qw & \qw & \qw & \qw & \qw & \qw & \qw & \rstick{\ket{\overline{Y_{m+n-1:n}}}} \\
		\lstick{\ket{0}} & \qw & \targ & \qw & \qw & \qw & \ctrl{3} & \ctrl{4} & \qw & \qw & \qw & \qw & \qw & \qw & \qw & \qw & \qw& \rstick{\ket{\overline{X_{m+n-1:n}  }}}\\
		\lstick{\ket{0}} & \qw & \qw & \qw &\gate{x} & \qw & \qw & \qw & \qw & \qw & \qw & \qw &\ctrl{4} & \qw & \qw & \qw & \qw &\rstick{\ket{Garbage} }\\
		\lstick{\ket{0}} & \qw & \qw & \qw &\qw &\gate{x} & \qw & \qw & \qw & \qw & \qw &\ctrl{4} & \qw & \qw & \qw & \qw & \qw &\rstick{\ket{Garbage} }\\
		\lstick{\ket{0}} & \qw & \qw & \qw & \qw & \qw &\gate{x} & \qw & \qw & \qw &\ctrl{4} & \qw & \qw & \qw & \qw & \qw & \qw &\rstick{\ket{Garbage} }\\
		\lstick{\ket{0}} & \qw & \qw & \qw & \qw & \qw & \qw &\gate{x} & \qw &\ctrl{4} & \qw & \qw & \qw & \qw & \qw & \qw & \qw &\rstick{\ket{Garbage} }\\
		\lstick{\ket{C_{Y,X}}} & \qw & \qw & \qw & \qw & \qw & \qw & \qw & \qw & \qw & \qw & \qw &\ctrl{7} & \qw & \qw & \qw & \qw& \rstick{\ket{C_{Y,X}}}\\
		\lstick{\ket{C_{Y+1,X}}} & \qw & \qw & \qw & \qw & \qw & \qw & \qw & \qw & \qw & \qw &\ctrl{5} & \qw & \qw & \qw & \qw  & \qw& \rstick{\ket{C_{Y+1,X}}}\\
		\lstick{\ket{C_{Y,X+1}}} & \qw & \qw & \qw & \qw & \qw & \qw & \qw & \qw & \qw & \ctrl{3} & \qw & \qw & \qw & \qw & \qw  & \qw& \rstick{\ket{C_{Y,X+1}}}\\
		\lstick{\ket{C_{Y+1,X+1}}} & \qw & \qw & \qw & \qw & \qw & \qw & \qw & \qw & \ctrl{1} & \qw  & \qw & \qw & \qw & \qw & \qw & \qw& \rstick{\ket{C_{Y+1,X+1}}}\\
		\lstick{\ket{0}} & \qw & \qw & \qw & \qw & \qw & \qw & \qw & \qw & \gate{x} & \qw & \qw & \qw & \qw & \qw &\ctrl{3} & \qw& \rstick{\ket{Garbage} }\\
		\lstick{\ket{0}} & \qw & \qw & \qw & \qw & \qw & \qw & \qw & \qw & \qw &\gate{x} & \qw & \qw & \qw &\ctrl{2} & \qw & \qw &\rstick{\ket{Garbage} }\\
		\lstick{\ket{0}} & \qw & \qw & \qw & \qw & \qw & \qw & \qw & \qw & \qw &\qw & \gate{x} & \qw &\ctrl{1} & \qw & \qw & \qw &\rstick{\ket{Garbage} }\\
		\lstick{\ket{0}} & \qw & \qw & \qw & \qw & \qw & \qw & \qw & \qw & \qw & \qw &\qw & \multigate{1}{x} & \multigate{1}{+} & \multigate{1}{+} & \multigate{1}{+} & \qw &\rstick{\ket{C_{\overline{Y},\overline{X}}}}\\
		\lstick{\ket{0}} & \qw & \qw & \qw & \qw & \qw & \qw & \qw & \qw & \qw & \qw &\qw & \ghost{x} & \ghost{+} & \ghost{+} & \ghost{+} & \qw &\rstick{\ket{Garbage} }\\
	}
	\]
	\caption{Proposed quantum bilinear interpolation circuit for the scale up operation.  The image is scaled up by a value $n$.  The notation $m+n-1:n$ means quantum register locations $m+n-1$ through $n$ for the output image position registers $\ket{\overline{Y}}$ and $\ket{\overline{X}}$.  The notation $n-1:0$ means quantum register locations $n-1$ through $0$ for the output image position registers $\ket{\overline{Y}}$ and $\ket{\overline{X}}$.}
	\label{bilinear-scale-up-image}
\end{figure}
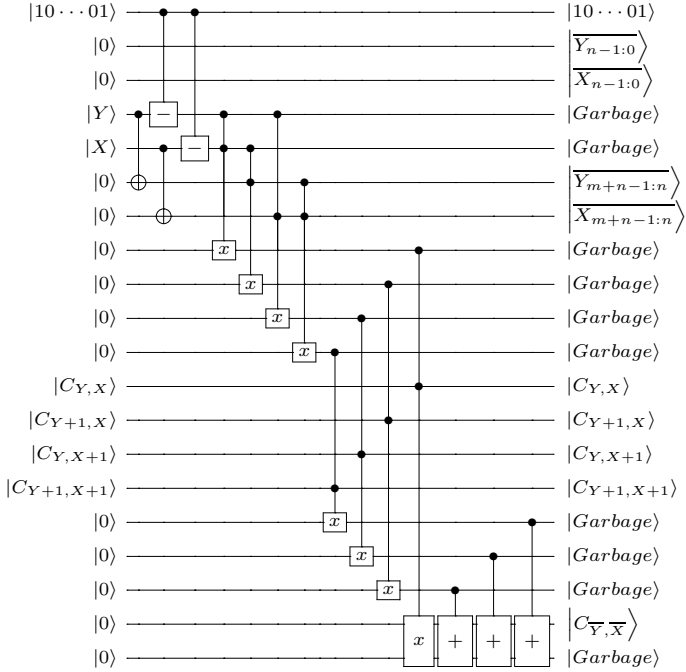

The proposed quantum circuit for bilinear interpolation is shown in Figure \ref{bilinear-scale-up-image} for the case of a scaling up an image by an integer value $n$.  Scaling up an image by an integer value $n$ results in increasing the original $y$ and $x$ positions of each pixel by $2^n$.  Inputs to the proposed quantum circuit is the original pixel positions $y$ and $x$ stored in $m$ bit quantum registers $\ket{Y}, \ket{X}$, respectively, along with the corresponding original pixel color information $C_{Y,X}$ stored in quantum register $\ket{C_{Y,X}}$.  Inputs also include the pixel color information for adjacent original pixels at locations ($y+1,x$), ($y,x+1$) and ($y+1,x+1$).  The color information for these adjacent pixels are stored in quantum registers $\ket{C_{Y+1,X}}$, $\ket{C_{Y,X+1}}$ and $\ket{C_{Y+1,X+1}}$ respectively.  At the end of computation, the circuitry returns the position information for the scaled pixel in quantum registers $\ket{\overline{Y}}$ and $\ket{\overline{X}}$ as well as the corresponding color of the scaled pixel in quantum register $\ket{C_{\overline{Y},\overline{X}}}$.

To scale up an image by a value $n$, the position value is multiplied by $2^n$.  By concatenating $n$ ancillae set to $0$ to quantum registers $\ket{X}$ and $\ket{Y}$ such that they have the values $X_{m-1} \cdots X_0 0 \cdots 0$ and $Y_{m-1} \cdots Y_0 0 \cdots 0$ we eliminate the need to use a quantum multiplication circuit.   These combined registers now contain the position information $\overline{Y}$ and $\overline{X}$, respectively, for the scaled up pixel.

To calculate the color information for the scaled pixel, our proposed quantum circuit must calculate equation \ref{blabs:1again}.  

{\small 
\begin{equation}
\left[ \begin{matrix}
\left(2^m - \frac{\overline{Y}}{2^n}\right) \cdot \left(2^m - \frac{\overline{X}}{2^n}\right) \cdot C_{Y,X} + \\
\left(\frac{\overline{Y}}{2^n}\right) \cdot (2^m -\left(\frac{\overline{X}}{2^n}\right) \cdot C_{Y+1,X} + \\
\left(2^m - \frac{\overline{Y}}{2^n}\right) \cdot \left(\frac{\overline{X}}{2^n}\right) \cdot C_{Y,X+1} + \\
\left(\frac{\overline{Y}}{2^n}\right) \cdot \left(\frac{\overline{X}}{2^n}\right) \cdot C_{Y+1,X+1} \\
\end{matrix} \right] \div 2^{2 \cdot m}
\label{blabs:1again}
\end{equation}
}

Where, $\frac{\overline{Y}}{2^n}$ corresponds to locations $\ket{\overline{Y_{m+n-1}}}$ through $\ket{\overline{Y_{n-1}}}$ of the combined position register $\ket{\overline{Y}}$ and $(\overline{X} - X \cdot 2^n))$ corresponds to locations $\ket{\overline{X_{m+n-1}}}$ through $\ket{\overline{X_{n-1}}}$ of the combined position register $\ket{\overline{X}}$ \cite{Zhou2017bilinear}.  The values at locations $\ket{\overline{Y_{m+n-1}}}$ through $\ket{\overline{Y_{n-1}}}$ and at locations $\ket{\overline{X_{m+n-1}}}$ through $\ket{\overline{X_{n-1}}}$ correspond to the original $y$ and $x$ positions of the pixel.    

To perform this computation, the proposed bilinear interpolation circuit requires a quantum multiplication circuit, a quantum addition circuit, and a quantum subtraction circuit.  Our circuit computes equation \ref{blabs:1again} by executing an algorithm identical to the proposed scale down circuit.  The algorithm our quantum bilinear interpolation circuit for the scale up operation implements is as follows:

\begin{itemize}

\item Step 1: Copy $\frac{\overline{Y}}{2^n}$ and $\frac{\overline{X}}{2^n}$ to ancillae with CNOT gates.

\item Step 2: Calculate $2^m - \frac{\overline{Y}}{2^n}$ and $2^m - \frac{\overline{Y}}{2^n}$ with quantum subtraction circuits.  

\item Step 3: Calculate the products $\left(2^m - \frac{\overline{Y}}{2^n}\right) \cdot \left(2^m - \frac{\overline{X}}{2^n}\right)$, $\left(\frac{\overline{Y}}{2^n}\right) \cdot \left(2^m - \frac{\overline{X}}{2^n}\right)$, $\left(2^m - \frac{\overline{Y}}{2^n}\right)) \cdot \left(\frac{\overline{X}}{2^n}\right)$ and $\left(\frac{\overline{Y}}{2^n}\right) \cdot \left(\frac{\overline{X}}{2^n}\right)$ with quantum multiplication circuits.

\item Step 4: Calculate the product terms $\left(2^m - \frac{\overline{Y}}{2^n}\right) \cdot \left(2^m - \frac{\overline{X}}{2^n}\right) \cdot C_{Y,X}$, $\left(\frac{\overline{Y}}{2^n}\right) \cdot (2^m -\left(\frac{\overline{X}}{2^n}\right) \cdot C_{Y+1,X}$, $\left(2^m - \frac{\overline{Y}}{2^n}\right) \cdot \left(\frac{\overline{X}}{2^n}\right) \cdot C_{Y,X+1}$ and $\left(\frac{\overline{Y}}{2^n}\right) \cdot \left(\frac{\overline{X}}{2^n}\right) \cdot C_{Y+1,X+1}$ with quantum multiplication circuits.

\item Step 5: Complete the calculation of equation \ref{blabs:1again} with quantum addition circuits. 

\end {itemize}

The values at locations $2 \cdot m - 1$ through $0$ of the quantum register with the result of Step 5 are not a part of the new pixel's color information $\left( C_{\overline{Y},\overline{X}} \right) $.  By not assigning these locations to the quantum register containing the new pixel's color information $\ket{C_{\overline{Y},\overline{X}}}$ we eliminate the need to use quantum division circuits.  The remaining locations are the new pixel's color information $\left( C_{\overline{Y},\overline{X}} \right) $ and will be assigned to the quantum register containing the new pixel's color information $\ket{C_{\overline{Y},\overline{X}}}$.

\subsection{Cost Analysis of the Proposed Bilinear Interpolation Circuit for the Scale Up Operation}
\label{cost-bi-up}

\begin{table}[tbhp]
\centering
\caption{Comparison of the Bilinear Interpolation Circuits for the Scale Up Operation in Terms of Number of Functional Blocks Used}
\label{bilinear-table:5}
\begin{tabular}{lccc}
\\ \midrule
Component & & 1 & proposed \\ \toprule
Adder & &$3$ & $3$ \\
Subtractor & & $4$ & $2$ \\
Divider & & $2$ & $0$* \\
Multiplier & & $8 $ & $8$ \\ \bottomrule
\multicolumn{4}{l}{1 is the design in \cite{Zhou2017bilinear}} \\
\multicolumn{4}{l}{* Our proposed design does not require a divider.} \\ \midrule
\end{tabular}
\end{table}

\begin{table*}[tbhp]
\centering
\caption{T Gate Comparison of Bilinear Interpolation Circuits for the Scale Up Operation}
\label{bilinear-table:4}
\begin{tabular}{lccc}
\\ \midrule
&  & T-count & \% impr. w.r.t. 1 \\ \toprule
1 & & $856 \cdot n^2 + 196 \cdot n - 98 + 8 \cdot \sum_{i = 1}^{log_2(n)} \frac{n}{2^i} \cdot \left( 14 \cdot \left( n + i - 2^{i-1} \right) -14 \right)$ & - \\ 
Proposed & & $64 \cdot n^2 - 12 \cdot n - 8$ & $ \approx 92.52 \%$ \\ \bottomrule
\multicolumn{4}{l}{1 is the design in \cite{Zhou2017bilinear}} \\
\end{tabular}
\end{table*}

Table \ref{bilinear-table:5} shows the comparison between the proposed quantum circuit for bilinear interpolation and the existing work in terms of total number of arithmetic operations.  Through careful layout of functional blocks in our proposed quantum circuit for bilinear interpolation we remove 2 quantum subtraction circuits from our design.  Further, by not assigning locations $2 \cdot m - 1$ through $0$ of the quantum register containing the result of equation \ref{blabs:1again} to the quantum register containing the new pixel's color information $\ket{C_{\overline{Y},\overline{X}}}$ we eliminate the need to use quantum division circuits.     

 Comparison of the T-count between the proposed quantum bilinear interpolation circuit for the scale up operation and the existing work in \cite{Zhou2017bilinear} is shown in Table \ref{bilinear-table:4}.  The design in \cite{Zhou2017bilinear} has a T-count of order $ \approx \mathcal{O}(n^2)$. The proposed design's T-count is of order $\mathcal{O}(n^2)$.  To calculate the T-count of the existing design in \cite{Zhou2017bilinear}, we determined the T-counts for the quantum circuits used in the design.  We determined that to calculate equation \ref{blabs:1}, the design in \cite{Zhou2017bilinear} uses a quantum addition circuit with a T-count of $28 \cdot n - 14$, a quantum subtraction circuit with a T-count of $28 \cdot n - 14$, a quantum multiplication circuit with a T-count of $7 \cdot n^2 + \sum_{i = 1}^{log_2(n)} \frac{n}{2^i} \cdot \left( 14 \cdot \left( n + i - 2^{i-1} \right) -14 \right) $, $8 \cdot n^2 - 4 \cdot n$ and a quantum division circuit with a T-count of $\approx 400 \cdot n^2$.  We compute the T-count for the proposed work and the existing design by multiplying the T-count for each quantum functional block by the number of times it is used and then sum the result.

Table \ref{bilinear-table:4} shows that the proposed quantum circuit for bilinear interpolation has an improvement ratio of $ \approx 92.52 \%$ in terms of T-count when performing the scale up operation.  Our proposed quantum bilinear interpolation circuit reduces T-count by (i) using T gate efficient functional blocks and (ii) reducing the number of required functional blocks to calculate the new pixel's color information (see equation \ref{blabs:1}).  Our proposed designs are based on the quantum arithmetic circuits illustrated in Section \ref{bi-components}.  As a result, our proposed bilinear interpolation quantum circuit is based on a quantum addition circuit with a T-count of $4 \cdot n$, a novel quantum subtraction circuit with a T-count of $4 \cdot n-4$ and a quantum multiplication circuit with a T-count of $8 \cdot n^2 - 4 \cdot n$. 
\section{Conclusion}

In this work, we presented T-count efficient quantum circuits for bilinear interpolation.  Proposed quantum circuits for the scale up operation and scale down operation are illustrated.  The building blocks used in our proposed quantum bilinear interpolation circuits (quantum adder, proposed quantum subtractor and quantum multiplication circuit) are also shown.  The proposed quantum bilinear interpolation circuits are compared and achieve significant T-count savings compared to the existing work.  They are also shown to require fewer quantum arithmetic circuits compared to the existing work.  We conclude that the proposed quantum bilinear interpolation circuits can be integrated in a larger quantum image processing circuit for NEQR encoded images where T-count is of primary concern.

\bibliographystyle{IEEEtran}
\bibliography{bilinear.bib}

\begin{thebibliography}{10}
\providecommand{\url}[1]{#1}
\csname url@samestyle\endcsname
\providecommand{\newblock}{\relax}
\providecommand{\bibinfo}[2]{#2}
\providecommand{\BIBentrySTDinterwordspacing}{\spaceskip=0pt\relax}
\providecommand{\BIBentryALTinterwordstretchfactor}{4}
\providecommand{\BIBentryALTinterwordspacing}{\spaceskip=\fontdimen2\font plus
\BIBentryALTinterwordstretchfactor\fontdimen3\font minus
  \fontdimen4\font\relax}
\providecommand{\BIBforeignlanguage}[2]{{%
\expandafter\ifx\csname l@#1\endcsname\relax
\typeout{** WARNING: IEEEtran.bst: No hyphenation pattern has been}%
\typeout{** loaded for the language `#1'. Using the pattern for}%
\typeout{** the default language instead.}%
\else
\language=\csname l@#1\endcsname
\fi
#2}}
\providecommand{\BIBdecl}{\relax}
\BIBdecl

\bibitem{Beach2004motivation}
G.~Beach, C.~Lomont, and C.~Cohen, ``Quantum image processing (quip),'' vol.
  2003-.\hskip 1em plus 0.5em minus 0.4em\relax Institute of Electrical and
  Electronics Engineers Inc., 2004, pp. 39--44.

\bibitem{Caraiman2012motivation}
S.~Caraiman and V.~Manta, ``\BIBforeignlanguage{eng}{Image processing using
  quantum computing},'' in \emph{\BIBforeignlanguage{eng}{System Theory,
  Control and Computing (ICSTCC), 2012 16th International Conference
  on}}.\hskip 1em plus 0.5em minus 0.4em\relax IEEE, October 2012, pp. 1--6.

\bibitem{Le2011FRQI}
\BIBentryALTinterwordspacing
P.~Q. Le, F.~Dong, and K.~Hirota, ``A flexible representation of quantum images
  for polynomial preparation, image compression, and processing operations,''
  \emph{Quantum Information Processing}, vol.~10, no.~1, pp. 63--84, Feb 2011.
  [Online]. Available: \url{https://doi.org/10.1007/s11128-010-0177-y}
\BIBentrySTDinterwordspacing

\bibitem{Zhang2013NEQR}
\BIBentryALTinterwordspacing
Y.~Zhang, K.~Lu, Y.~Gao, and M.~Wang, ``Neqr: a novel enhanced quantum
  representation of digital images,'' \emph{Quantum Information Processing},
  vol.~12, no.~8, pp. 2833--2860, Aug 2013. [Online]. Available:
  \url{https://doi.org/10.1007/s11128-013-0567-z}
\BIBentrySTDinterwordspacing

\bibitem{Zhou2017bilinear}
R.-G. Zhou, W.~Hu, P.~Fan, and H.~Ian, ``Quantum realization of the bilinear
  interpolation method for {NEQR},'' \emph{Scientific Reports}, vol.~7, no.~1,
  2017.

\bibitem{Jiang2015interpolation}
N.~Jiang and L.~Wang, ``\BIBforeignlanguage{eng}{Quantum image scaling using
  nearest neighbor interpolation},'' \emph{\BIBforeignlanguage{eng}{Quantum
  Information Processing}}, vol.~14, no.~5, pp. 1559--1571, 2015.

\bibitem{YanFei2017rotation}
\BIBentryALTinterwordspacing
F.~Yan, K.~Chen, S.~Venegas-Andraca, and J.~Zhao,
  ``\BIBforeignlanguage{eng}{Quantum image rotation by an arbitrary angle},''
  \emph{\BIBforeignlanguage{eng}{Quantum Information Processing}}, vol.~16,
  no.~11, pp. 1--20, 2017. [Online]. Available:
  \url{http://search.proquest.com/docview/1950084195/?pq-origsite=primo}
\BIBentrySTDinterwordspacing

\bibitem{Maslov2013cliffordT}
M.~Amy, D.~Maslov, M.~Mosca, and M.~Roetteler, ``A meet-in-the-middle algorithm
  for fast synthesis of depth-optimal quantum circuits,'' \emph{IEEE
  Transactions on Computer-Aided Design of Integrated Circuits and Systems},
  vol.~32, no.~6, pp. 818--830, June 2013.

\bibitem{PalerIOP}
\BIBentryALTinterwordspacing
A.~Paler, I.~Polian, K.~Nemoto, and S.~J. Devitt, ``Fault-tolerant, high-level
  quantum circuits: form, compilation and description,'' \emph{Quantum Science
  and Technology}, vol.~2, no.~2, p. 025003, 2017. [Online]. Available:
  \url{http://stacks.iop.org/2058-9565/2/i=2/a=025003}
\BIBentrySTDinterwordspacing

\bibitem{HowardMark2017Tgates}
M.~Howard and E.~Campbell, ``\BIBforeignlanguage{eng}{Application of a resource
  theory for magic states to fault-tolerant quantum computing},''
  \emph{\BIBforeignlanguage{eng}{Physical review letters}}, vol. 118, no.~9,
  2017.

\bibitem{Zhou-T}
\BIBentryALTinterwordspacing
X.~Zhou, D.~W. Leung, and I.~L. Chuang, ``Methodology for quantum logic gate
  construction,'' \emph{Phys. Rev. A}, vol.~62, p. 052316, Oct 2000. [Online].
  Available: \url{https://link.aps.org/doi/10.1103/PhysRevA.62.052316}
\BIBentrySTDinterwordspacing

\bibitem{Kiteav2005faulttolerant}
\BIBentryALTinterwordspacing
S.~Bravyi and A.~Kitaev, ``Universal quantum computation with ideal clifford
  gates and noisy ancillas,'' \emph{Phys. Rev. A}, vol.~71, p. 022316, Feb
  2005. [Online]. Available:
  \url{https://link.aps.org/doi/10.1103/PhysRevA.71.022316}
\BIBentrySTDinterwordspacing

\bibitem{Cody2012toffoligatebuild}
\BIBentryALTinterwordspacing
C.~Jones, ``Low-overhead constructions for the fault-tolerant toffoli gate,''
  \emph{Phys. Rev. A}, vol.~87, p. 022328, Feb 2013. [Online]. Available:
  \url{https://link.aps.org/doi/10.1103/PhysRevA.87.022328}
\BIBentrySTDinterwordspacing

\bibitem{Gidney20184TgateToffolibuild}
\BIBentryALTinterwordspacing
C.~Gidney, ``Halving the cost of quantum addition,'' \emph{{Quantum}}, vol.~2,
  p.~74, Jun. 2018. [Online]. Available:
  \url{https://doi.org/10.22331/q-2018-06-18-74}
\BIBentrySTDinterwordspacing

\bibitem{Draper}
\BIBentryALTinterwordspacing
S.~A. {Cuccaro}, T.~G. {Draper}, S.~A. {Kutin}, and D.~{Petrie Moulton}, ``{A
  new quantum ripple-carry addition circuit},'' \emph{eprint
  arXiv:quant-ph/0410184}, Oct. 2004. [Online]. Available:
  \url{https://arxiv.org/abs/quant-ph/0410184}
\BIBentrySTDinterwordspacing

\bibitem{Thapliyal2016addsub}
H.~Thapliyal, ``Mapping of subtractor and adder-subtractor circuits on
  reversible quantum gates,'' in \emph{Transactions on Computational Science
  XXVII}.\hskip 1em plus 0.5em minus 0.4em\relax Springer, 2016, pp. 10--34.

\bibitem{edgard2017multiplier}
E.~{Mu{\~n}oz-Coreas} and H.~{Thapliyal}, ``{T-count Optimized Design of
  Quantum Integer Multiplication},'' \emph{ArXiv: 1706.05113}, Jun. 2017.

\end{thebibliography}

\end{document}